\documentclass[twocolumn,showpacs,preprintnumbers,amsmath,amssymb]{revtex4}

\usepackage{graphicx}
\usepackage{dcolumn}
\usepackage{bm}
\begin{document}
\title{Umklapp scattering of pairs in BCS superconductivity theory}
\date{16 July 2004}
\author{X.\ H.\ Zheng and D.\ G.\ Walmsley}
\affiliation{Department of Pure and Applied Physics, The Queen's
University of Belfast, Belfast BT7 1NN, Northern Ireland}
\email{xhz@qub.ac.uk}

\begin{abstract}
The BCS theory of superconductivity is extended to recognize
pairing of electrons by both normal and umklapp phonon scattering.
Application of the variational approach shows that coexistence of
normal and umklapp scattering frustrates superconductivity.
\end{abstract}

 \maketitle

\section{Introduction}
\label{sec:Introduction} We wish to call attention to a result
which, to our knowledge, has not previously been noticed: in a BCS
(Bardeen, Cooper and Schrieffer) superconductor~\cite{BCS} a range
of pair end states may be accessed via both normal and umklapp
scattering.  The situation resembles a double-slit experiment in
which electrons have two paths to access the screen and cancel
each other periodically in the production of the analogue of
Young's fringes.

Umklapp scattering is compatible with the canonical
transformation~\cite{Frohlich} which is the basis of the BCS
theory. Discussion about umklapp contributions to
superconductivity started almost immediately after the BCS theory
was forged~\cite{Pines,Morel}.  This discussion has gathered
renewed momentum after the high $T_c$ superconductors have been
discovered~\cite{Cannon,Anderson,Fukukawa,Walker,Chuang,Armitage}.
It is therefore timely to discuss our new result in some detail.

We show that the trial electron wave function in the standard BCS
theory is not designed to accommodate states with multiple access
paths. We find that, if we allow all the end states of both normal
and umklapp scattering to be generated by this wave function, then
we are not able to normalize it.

We extend the approach of BCS, who already modified their $T = 0$
trial function to accommodate single particles and excited pairs
that appear at $T > 0$.  We modify the BCS trial function at both
$T = 0$ and $T > 0$ to include occupation of states by both normal
and umklapp scattering. We show that the modified trial function
is always normalized.

We also show that the effect of umklapp scattering is to cancel
the pairing effect of normal electron-phonon scattering.  The
cancellation is actually the physical foundation for the average
phonon frequency, significantly lower than the Debye frequency,
introduced by BCS. We restrict our discussion exclusively to the
BCS formalism, based on the principle of variation. We do not
discuss for example the Eliashberg formalism~\cite{Eliashberg}
which derives from a different principle.

\section{Hamiltonian}
\label{sec:Hamiltonian} In the BCS theory the standard procedure
of variation is followed. It starts with the reduced Hamiltonian
\begin{equation}
H_{\hbox{\scriptsize BCS}} = 2\sum_{\bf k}\epsilon_{\bf k}b^+_{\bf
k}b_{\bf k} -\sum_{\bf k,k'}V_{\bf k, k'}b^+_{\bf k'}b_{\bf
k}\label{eq:H_BCS}
\end{equation}
where $V_{\bf k, k'}$ measures the strength of the electron-phonon
interaction in second order,
\begin{equation}
b^+_{\bf k} = a^+_{{\bf -k} \downarrow}a^+_{{\bf k} \uparrow}
\label{eq:b}
\end{equation}
generates pairs, $a^+$ being the particle generation operator,
$\bf k$ electron wave vector, $\uparrow$ and $\downarrow$ spin. In
Eq.~(\ref{eq:H_BCS}) $\bf k$ runs over each and every state of the
first electron Brillouin zone exactly once with spin $\uparrow$
(we assume a single band throughout our discussion).
Simultaneously $\bf-k$ runs over the same range with spin
$\downarrow$. An alternative approach, which we do not adopt, is
to let $\bf k$ in Eq.~(\ref{eq:H_BCS}) run over half the Brillouin
zone and sum the spins. Either way, we may have spin $\uparrow$
and $\downarrow$ for any specific value of $\bf k$.  Consequently
the paired particles may be in $({\bf k}\uparrow,{\bf
-k}\downarrow)$ or $({\bf k}\downarrow,{\bf -k}\uparrow)$.

In Eq.~(\ref{eq:H_BCS}) the electron term is in $b^+_{\bf k}$ and
$b_{\bf k}$, which is not entirely equivalent to the standard
electron Hamiltonian
\begin{equation}
\sum_{\bf k}\left(\epsilon_{\bf k}a^+_{{\bf k}\uparrow}a_{{\bf
k}\uparrow} + \epsilon_{\bf k}a^+_{{\bf k}\downarrow}a_{{\bf
k}\downarrow}\right) \label{eq:H_e}
\end{equation}
The electron term in Eq.~(\ref{eq:H_BCS}) will vanish when applied
to single particles, but the Hamiltonian in
Expression~(\ref{eq:H_e}) will not: Eq.~(\ref{eq:H_BCS}) is
designed only for a pair ensemble. Indeed even
Expression~(\ref{eq:H_e}) is designed only for electrons in
crystals, where we have discrete electron energy $\epsilon_{\bf
k}$, which is produced automatically by the Hamiltonian when an
electron is detected by the particle destruction operator $a_{\bf
k\uparrow}$ or $a_{\bf k\downarrow}$.

\section{Trial function}
\label{sec:Trial function} Next is introduced the trial wave
function for quasi-particles. At $T = 0$ this function is of the
following form in the BCS theory~\cite{BCS}:
\begin{equation}
|\Psi\rangle = \prod _{\bf k}\big(\sqrt{1-h_{\bf k}} +
\sqrt{h_{\bf k}}\;b^+_{\bf k}\big)|0\rangle \label{eq:ground}
\end{equation}
$h_{\bf k}$ is the pair occupation probability and $|0\rangle$ the
pair vacuum.  Here we seek the equilibrium configuration of the
particle ensemble under the influence of the electron-phonon
interaction. The trial function must accommodate all the initial
and end states of the paired quasi-particles, which are indicated
by $\bf k$ and $\bf k'$ in Eq.~(\ref{eq:H_BCS}), with occupancy
probabilities $h_{\bf k}$ to be determined for both the initial
and end states in a single step of the process of variation.

Now in Eq.~(\ref{eq:ground}) the product over $\bf k$ also runs
over the first electron Brillouin zone, i.e.\ the number of orbits
in Eq.~(\ref{eq:ground}) equals the number of Bloch states.
However, this does not mean that we cannot add further orbits to
Eq.~(\ref{eq:ground}). The growth of the superconductive energy
gap can be compared with the splitting of the spectral line of say
the hydrogen atom, where the electron has several configurations,
with different patterns of particle location probability and/or
spin but the same energy.  This energy degeneracy splits to reveal
further structure in the presence of e.g.\ an external magnetic
field. Similarly, a particle pair has two configurations,
generated by
\begin{equation}
\sqrt{1-h_{\bf k}} + \sqrt{h_{\bf k}}\;b^+_{\bf k}
\label{eq:normal}
\end{equation}
and
\begin{equation}
\sqrt{1-h_{\bf k}}\;b^+_{\bf k} - \sqrt{h_{\bf k}}
\label{eq:umklapp}
\end{equation}
respectively, whose energy splits from the Bloch energy
$\epsilon_{\bf k}$ in the presence of the electron-phonon
interaction. The energy of the pair generated by
Expression~(\ref{eq:normal}) shifts downwards to give the energy
gap. The energy of the pair generated by
Expression~(\ref{eq:umklapp}) shifts upwards. Apparently, in
addition to the Bloch orbits, we may have two more orbits per
primitive cell. Eq.~(\ref{eq:ground}) does not include all the
available orbits, because the paired states there suffice to
accommodate the physics: the superconductive current is carried
only by the paired quasi-particles with energy shifted downwards
under the electron-phonon interaction.

When $T > 0$, Eq.~(\ref{eq:ground}) no longer suffices to
accommodate the physics.  Indeed, if we continue to employ
Eq.~(\ref{eq:ground}) as our trial function, then we have no
chance to see what happens when superconductive pairs break into
single particles or are excited under the attack of thermal
phonons. To cope, BCS modified Eq.~(\ref{eq:ground}) into the
following form:
\begin{eqnarray}
&& |\Psi\rangle_{\mbox{exc}} = \prod _{{\bf k''}(\mathcal
S)}a^+_{{\bf k''}\sigma}\prod _{{\bf k'}(\mathcal
P)}\big(\sqrt{1-h_{\bf k'}}\;b^+_{\bf k'}\nonumber\\
&& \;\;\;\;\;\;- \sqrt{h_{\bf k'}}\big)\prod _{{\bf k}(\mathcal
G)}\big(\sqrt{1-h_{\bf k}} + \sqrt{h_{\bf k}}\;b^+_{\bf
k}\big)|0\rangle \label{eq:excited}
\end{eqnarray}
where $\mathcal S$, $\mathcal P$ and $\mathcal G$ specify the
states occupied by single particles, excited pairs and ground
pairs respectively, $\sigma$ (could be denoted as $\sigma_{\bf
k}$) stands for either $\uparrow$ or $\downarrow$ depending the
value of $\bf k$~\cite{BCS}.  It is interesting to note that in
Eq.~(\ref{eq:excited}) ${\bf k}$, ${\bf k}'$ and ${\bf k}''$ do
not have to be different. Indeed we can check through direct
calculation that, when $\bf k = k' = k''$, the wave functions of
the single particle, the excited pair and the ground pair are
orthogonal to each other: the total number of states in
Eq.~(\ref{eq:excited}) can be three times as large as that in
Eq.~(\ref{eq:ground}).

\begin{figure}
\resizebox{8cm}{!}{\includegraphics{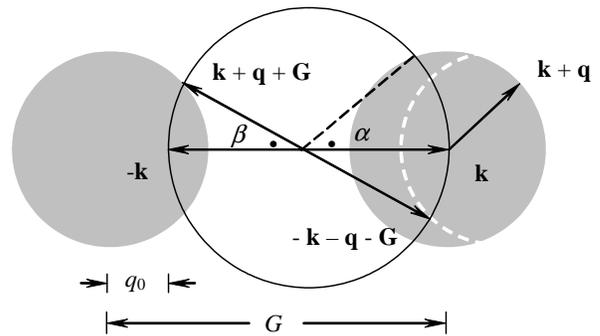}}
\caption{\label{fig:fig1} A spherical Fermi surface (open circle)
and a pair of electrons in states $\bf k$ and $\bf -k$. This pair
is scattered into states $\bf k + q + G$ and $\bf -k - q - G$ in
the umklapp process, $\bf q$ being the phonon wave vector, $\bf G$
the reciprocal lattice vector. The shaded circles represent two
identical spherical phonon zones. We use angles $\alpha$ and
$\beta$ to measure the size of the sections of the Fermi surface
being intersected by these phonon zones.  We have $\alpha > \beta$
because a larger section of the Fermi surface is intersected by
the phonon sphere centered at this surface (the other sphere is
centered above the Fermi surface at height $q_0$). As a result,
the electron state $\bf -k - q - G$ must lie inside the phonon
sphere centered at $\bf k$, where all the states are involved in
normal scattering. When $\bf k + q + G$ runs over the shaded
phonon zone on the left, $\bf -k - q - G$ also runs over a sphere
indicated in part by the broken white circle (centered above the
Fermi surface at height $q_0$ to mirror the phonon sphere on the
left).}
\end{figure}

\section{Destination state with multiple access paths}
\label{sec:Umklapp scattering} Consider in Fig.~\ref{fig:fig1} a
spherical Fermi surface that does not intersect the boundary of a
Brillouin zone: suppose also there is a minimum separation $q_0$
between adjacent Fermi surfaces. An electron in state $\bf k$ may
be scattered into $\bf k + q$ to lie on the adjacent Fermi surface
if the phonon momentum $q$ exceeds $q_0$. Since it makes no
physical difference if the reciprocal lattice vector $\bf G$
(parallel to $\bf k$ for simplicity) is added to an electron
state, we can replace $\bf k + q$ with $\bf k + q + G$, which lies
on the original Fermi surface. This state is known as the end
state of umklapp scattering and, in a non-superconducting metal,
cannot be accessed through normal scattering (in the reduced zone
scheme), at least for a range of $\bf q$ values.

In a superconductor one destination state may have multiple access
paths. For example in Fig.~\ref{fig:fig1} $\bf k$ and $\bf -k$ are
paired and, correspondingly, $\bf k + q + G$ and $\bf -k - q - G$
are also paired: scattering of the particle pair is synchronized,
reminiscent of the synchronized motion of rotating doors.  In the
umklapp process in Fig.~\ref{fig:fig1} we have the following
possible chains of events concerning state occupation:
\begin{equation}
{\bf k}\rightarrow{\bf k + q}\rightarrow{\bf k + q +
G}\rightarrow{\bf - k - q - G}\label{eq:path1}
\end{equation}
The first two arrows represent scattering and umklapp folding. The
third arrow denotes synchronized occupation of the state $\bf -k -
q - G$ by the particle initially in state $\bf -k$.  These events
can be compared with say passage of an electron through a slit, or
reflection of a photon by a mirror, in an interference experiment.
It is evident from Fig.~\ref{fig:fig1} that $\bf -k - q - G$ lies
inside the phonon sphere centered at $\bf k$, where all the
electron normal-scattering end states lie, and therefore we also
have the following normal-scattering event:
\begin{equation}
{\bf k}\rightarrow{\bf -k - q - G}\label{eq:path2}
\end{equation}
Apparently the state $\bf -k - q - G$ can be accessed via two
different paths.

\begin{figure} \resizebox{8cm}{!}{\includegraphics{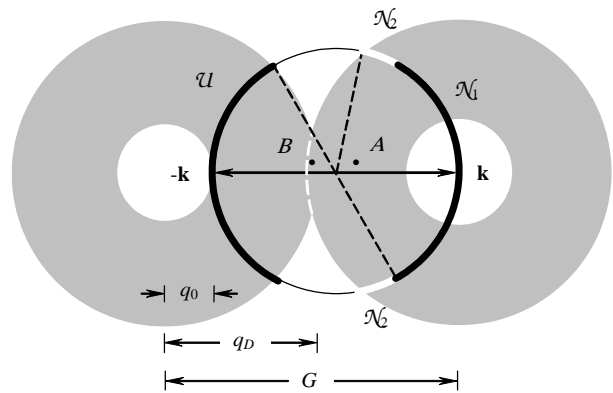}}
\caption{\label{fig:fig2} A spherical Fermi surface (open circle)
with two phonon spheres of radius $q_D$ (Debye momentum).  In
umklapp scattering the particle in $\bf k$ is scattered into
$\mathcal U$, which is a layer about the Fermi surface shown as a
thick black arc. In normal scattering the particle in state $\bf
k$ is scattered into $\mathcal N = \mathcal N_1 + \mathcal N_2$,
which is also a layer about the Fermi surface shown as a thick arc
in black and white, $\mathcal N_1$ and $\mathcal U$ are symmetric
with respect to the center of the Fermi sphere. The sizes of
$\mathcal N$ and $\mathcal U$ are measured by $A$ and $B$, which
are maximum values of $\alpha$ and $\beta$ in Fig.~\ref{fig:fig1}.
There is a hole in the phonon spheres to indicate that umklapp
scattering is not invoked when the phonon momentum is less than
$q_0$.}
\end{figure}

We use $\mathcal N$ and $\mathcal U$ to denote collections of all
the (single) end states of normal and umklapp scattering,
respectively, which form thin layers about the Fermi surface, in
which empty states are available to accommodate scattered
particles. In Fig.~\ref{fig:fig2} $\mathcal N$ is a section of the
Fermi surface intersected by the phonon sphere centered at $\bf
k$, whose radius is $q_D$ (Debye momentum) to include all the
phonons allowed in the Debye model. This phonon sphere may be
transported to the left by a lattice constant $G$. The transported
phonon sphere also intersects a section, $\mathcal U$, of the
Fermi surface to accommodate particles scattered from $\bf k$ in
the umklapp process. We separate $\mathcal N$ into two sub-sets,
$\mathcal N_1$ and $\mathcal N_2$, where $\mathcal N_1$
accommodates particles scattered from $\bf-k$ in the umklapp
process.  It is apparent from Fig.~\ref{fig:fig2} that an
extensive range of states in $\mathcal N$ can be accessed via both
normal and umklapp scattering.

We also use $\mathcal N$ and $\mathcal U$ to specify paired end
states of normal and umklapp scattering (or normal and umklapp
pairs for short).  In Fig.~\ref{fig:fig2}) the single particle in
state $\bf k$ may have spin $\uparrow$ or $\downarrow$.
Consequently the normal or umklapp pair may be in $(\bf
k'\uparrow, -k'\downarrow)$ or $(\bf k'\downarrow, k'\uparrow)$
after scattering.  The shape of $\mathcal N$ and $\mathcal U$ is
the same in both cases.

\section{Limitation of BCS trial function}
\label{sec:Limitation} The BCS trial function is not designed to
accommodate states with multiple access paths, such as those
described in Expressions (\ref{eq:path1}) and (\ref{eq:path2}). To
appreciate this, we study the example in Fig.~\ref{fig:fig2},
where we illustrate how states are linked by $V_{\bf k, k'}$ in
Eq.~(\ref{eq:H_BCS}) when $\bf k$ is in the direction of $\bf G$.
We write Eq.~(\ref{eq:ground}) in the following equivalent form
\begin{eqnarray}
|\Psi\rangle = \prod _{\bf k'(\mathcal N_1)}\prod _{\bf
k'(\mathcal N_2)}\prod _{\bf k'(\mathcal U)}\big(\sqrt{1-h_{\bf
k'}} + \sqrt{h_{\bf k'}}\;b^+_{\bf
k'}\big)|\Phi\rangle\label{eq:ground2}
\end{eqnarray}
with
\begin{equation}
|\Phi\rangle = \prod_{\bf k''(\mathcal R)}\big(\sqrt{1-h_{\bf
k''}} + \sqrt{h_{\bf k''}}\;b^+_{\bf k''}\big)|0\rangle\nonumber
\end{equation}
where $\mathcal R$ specifies states not in $\mathcal N_1$,
$\mathcal N_2$ and $\mathcal U$. By definition for each $\bf k'$
in $\mathcal U$ we have $\bf -k'$ in $\mathcal N_1$ and vice
versa, so that we can write the product over $\bf k'(\mathcal
N_1)$ in Eq.~(\ref{eq:ground2}) as
\begin{eqnarray}
\prod_{\bf k'(\mathcal U)}\big(\sqrt{1-h_{\bf -k'}} + \sqrt{h_{\bf
-k'}}\;b^+_{\bf -k'}\big)\label{eq:N2}
\end{eqnarray}
Combining Eqs.~(\ref{eq:ground2}) and (\ref{eq:N2}), we find
\begin{eqnarray}
\langle\Psi|\Psi\rangle = \prod_{\bf k'(\mathcal{U})}\Big[1
+2\sqrt{h_{\bf k'}(1 - h_{\bf k'})}\nonumber\\
\times\sqrt{h_{\bf -k'}(1 - h_{\bf
-k'})}\;\Big]\langle\Phi|\Phi\rangle \label{eq:psi3}
\end{eqnarray}
so that $\langle\Psi|\Psi\rangle > 1$ even when
$\langle\Phi|\Phi\rangle = 1$. Obviously the standard BCS theory
is not designed to endorse the umklapp process in
Fig.~\ref{fig:fig1}.

We cannot attribute the self-contradictory result in
Eq.~(\ref{eq:psi3}) to counting $\bf k + q$ and $\bf k + q + G$ in
Fig.~\ref{fig:fig1} as different states, because the states in
Eq.~(\ref{eq:ground2}) are all in the first electron Brillouin
zone (Fig.~\ref{fig:fig2}).  We might like to attribute
Eq.~(\ref{eq:psi3}) to the proposition that normal and umklapp
scattering drive two pairs into the same state, and thus violate
the exclusion principle. However, the entity scattered in both the
normal and umklapp processes is the same pair in $\bf k$ and $\bf
-k$ (Figs.~\ref{fig:fig1}, \ref{fig:fig2}): we do not have two
pairs to violate that principle. It appears that, in
Eq.~(\ref{eq:ground}), the number of orbits is simply not enough
to cope with the situation that some areas of the Fermi surface,
such as $\mathcal N_1$ in Fig.~\ref{fig:fig2}, suffer enhanced
high probability of pair bombardment when normal and umklapp
scattering coexist. Indeed Eq.~(\ref{eq:psi3}) does suggest that
the number of pairs landing in $\mathcal U$ and $\mathcal N_1$
exceeds the limit of its design.

\section{Modified trial function}
\label{sec:Modified trial function}We extend the approach of BCS,
who modified the trial function at $T = 0$, Eq.~(\ref{eq:ground}),
into the trial function at $T > 0$, Eq.~(\ref{eq:excited}), to
accommodate single particles and excited pairs. We will modify the
trial function at both $T = 0$ and $T > 0$ to accommodate umklapp
scattering. We will add more orbits to Eq.~(\ref{eq:ground}) to
allow umklapp pairs to land in say $\mathcal U$ and $\mathcal N_1$
in Fig.~\ref{fig:fig2}, without self-contradiction as in
Eq.~(\ref{eq:psi3}). Since Expression~(\ref{eq:normal}) has
already been used to generate normal pairs in
Eq.~(\ref{eq:ground}), we have no choice but to use Expression
(\ref{eq:umklapp}) to generate umklapp pairs and find
\begin{eqnarray}
|\Psi\rangle &=& \prod_{\bf k({\mathcal U})}\big(\sqrt{1-h_{\bf
k}}\;b^+_{\bf k}\nonumber\\
&-& \sqrt{h_{\bf k}}\big)\prod_{\bf k({\mathcal
N})}\big(\sqrt{1-h_{\bf k}} + \sqrt{h_{\bf k}}\;b^+_{\bf
k}\big)|0\rangle\label{eq:ground3}
\end{eqnarray}
as our trial electron wave function in the presence of umklapp
scattering.

According to Eq.~(\ref{eq:ground3}) the equilibrium state of the
particle ensemble consists of a mixture of normal and umklapp
pairs. We find through direct calculation that we always have
$\langle\Psi|\Psi\rangle = 1$ if we can prove
\begin{equation}
h_{\bf k} = h_{\bf -k}\label{eq:h}
\end{equation}
where $\bf k$ and $\bf -k$ are specified by $\mathcal U$ and
$\mathcal N$, respectively, in order to avoid a self-contradiction
similar to that in Eq.~(\ref{eq:psi3}). Eq.~(\ref{eq:h}) must be
true, because both the Fermi sea and reciprocal lattice are
symmetric with respect to ${\bf k} = 0$: interchanging $\bf k$ and
$\bf -k$ makes no difference in Eq.~(\ref{eq:H_BCS}), the
Hamiltonian.

We have a more serious problem: in Eq.~(\ref{eq:ground3}) we have
actually assumed that $h_{\bf k}$ depends only on the numerical
value of $\bf k$, regardless of whether $\bf k$ is specified by
$\mathcal N$ or $\mathcal U$. Since the condition to find $h_{\bf
k}$ may be different in $\mathcal N$ and $\mathcal U$, it is still
possible that Eq.~(\ref{eq:ground3}) could be self-contradictory,
as was Eq.~(\ref{eq:ground}), in the presence of umklapp
scattering.  We will address this in the next section.

\section{Occupancy of normal and umklapp pairs}
\label{sec:Occupancy} We use Eq.~(\ref{eq:ground3}) to evaluate
the energy of the pair ensemble
\begin{equation}
W = \langle\Psi|H_{\hbox{\scriptsize{BCS}}}|\Psi\rangle
\label{eq:W}
\end{equation}
where $H_{\hbox{\scriptsize{BCS}}}$ is defined in
Eq.~(\ref{eq:H_BCS}). We find
\begin{eqnarray}
W = 2\sum_{\bf k({\mathcal N})}\epsilon_{\bf k}h_{\bf k} +
2\sum_{\bf k({\mathcal U})}\epsilon_{\bf k}(1 - h_{\bf k})\nonumber\\
 -W_{11} - W_{22} + W_{12} + W_{21} \label{eq:W2}
\end{eqnarray}
with
\begin{eqnarray}
&&W_{11} = \sum_{\bf k(\mathcal N)}\sum_{\bf k'(\mathcal N)}V_{\bf
k, k'}\sqrt{h_{\bf k}(1 - h_{\bf k})}\sqrt{h_{\bf k'}(1
- h_{\bf k'})}\nonumber\\
&&W_{12} =\sum_{\bf k(\mathcal N)}\sum_{\bf k'(\mathcal U)}V_{\bf
k, k'}\sqrt{h_{\bf k}(1 - h_{\bf k})}\sqrt{h_{\bf k'}(1 - h_{\bf
k'})}\nonumber
\end{eqnarray}
$W_{21}$ and $W_{22}$ are identical to $W_{12}$ and $W_{11}$,
respectively, save that $\mathcal N$ is replaced by $\mathcal U$
and vice versa.

We minimize $W$ in Eq.~(\ref{eq:W2}) with respect to $h_{\bf k}$.
When $\bf k$ is specified by $\mathcal N$ we have
\begin{eqnarray}
\frac{\partial W}{\partial h_{\bf k}} &=& 2\epsilon_{\bf k} -
\frac{1 - 2h_{\bf k}}{\sqrt{h_{\bf k}(1 - h_{\bf k})}}\nonumber\\
&\times& \biggl[\;\sum_{\bf k'(\mathcal N)}-\sum_{\bf k'(\mathcal
U)}\;\biggr]V_{\bf k, k'}\sqrt{h_{\bf k'}(1 - h_{\bf
k'})}\label{eq:dW1}
\end{eqnarray}
Here $\epsilon_{\bf k}$ and $h_{\bf k}$ are linked together by the
constraint
\begin{equation}
h_{\bf k} = \frac{1}{2}\biggl[1 - \frac{\epsilon_{\bf k}}{E({\bf
k})}\biggr]\label{eq:constrain1}
\end{equation}
with $E({\bf k}) = \sqrt{\Delta^2({\bf k}) + \epsilon^2_{\bf k}}$.
Letting $\partial W/\partial h_{\bf k} = 0$, we find from
Eqs.~(\ref{eq:dW1}) and (\ref{eq:constrain1})
\begin{equation}
\Delta({\bf k})= \biggl[\;\sum_{{\bf k'}(\mathcal N)} -\sum_{{\bf
k'}(\mathcal U)}\;\biggr]V_{\bf k, k'}\frac{\Delta({\bf k'})}
{2E({\bf k'})} \label{eq:self}
\end{equation}
as our self-consistent equation, which is identical to the
standard BCS self-consistent equation~\cite{BCS}, save the terms
in $\bf k'(\mathcal U)$, which reduce and may even cancel the
effect of the terms in $\bf k'(\mathcal N)$.

When $\bf k$ is specified by $\mathcal U$, we have
\begin{eqnarray}
\frac{\partial W}{\partial h_{\bf k}} &=& -2\epsilon_{\bf k} +
\frac{1 - 2h_{\bf k}}{\sqrt{h_{\bf k}(1 - h_{\bf k})}}\nonumber\\
&\times& \biggl[\;\sum_{\bf k'(\mathcal N)}-\sum_{\bf k'(\mathcal
U)}\;\biggr]V_{\bf k, k'}\sqrt{h_{\bf k'}(1 - h_{\bf
k'})}\label{eq:dW2}
\end{eqnarray}
which also leads to Eq.~(\ref{eq:self}).  Apparently we have the
same $\Delta(\bf k)$ for the same numerical value of $\bf k$,
regardless of whether $\bf k$ is specified by $\mathcal N$ or
$\mathcal U$. This $\Delta(\bf k)$ leads through
Eq.~(\ref{eq:constrain1}) to $h_{\bf k}$, which also depends only
on the numerical value of $\bf k$, and thus ensures
$\langle\Psi|\Psi\rangle = 1$ for Eq.~(\ref{eq:ground3}).

\begin{figure} \resizebox{8cm}{!}{\includegraphics{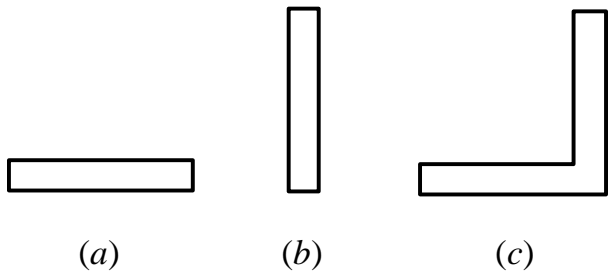}}
\caption{\label{fig:fig3} Configurations of (a) bar with minimum
energy, (b) bar with maximum energy and (c) two bar system.}
\end{figure}

\section{Frustration of superconductivity}
\label{sec:Frustration}The principle of variation can be compared
with shaking, which shapes a physical system into a configuration
with minimum energy, such as the horizontal orientation of the bar
in Fig.~\ref{fig:fig3} (a) in a gravitational field. However the
process of variation also produces a configuration with maximum
energy, such as the vertical orientation of the bar in
Fig.~\ref{fig:fig3} (b). Although we should choose the
configuration with minimum energy whenever possible, some physical
systems may frustrate us from doing so. For example, when the two
bars are joined together to become an `L' in Fig.~\ref{fig:fig3}
(c), we cannot minimize their energy as readily as we do in
Fig.~\ref{fig:fig3} (a), because minimum energy of one bar implies
maximum energy of the other, and vice versa.

We have a similar situation when normal and umklapp scattering
coexist.  The pair occupation probability, $h_{\bf k}$, has a
phase-like property, which becomes clear when we let $h_{\bf k} =
\cos^2\theta_{\bf k}$, so that the normal pair generation operator
in Expression (\ref{eq:normal}) becomes
\begin{equation}
\sin\theta_{\bf k} + \cos\theta_{\bf k}\;b^+_{\bf k}
\label{eq:normal-phase}
\end{equation}
The resultant wave function is normalized at any value of
$\theta_{\bf k}$.  However, in order to maximize the energy gap,
we have to let
\begin{equation}
\cos^2\theta_{\bf k} = \frac{1}{2}\biggl[1 - \frac{\epsilon_{\bf
k}}{E({\bf k})}\biggr]\label{eq:constrain1-phase}
\end{equation}
which is equivalent to Eq.~(\ref{eq:constrain1}). We can also
write the umklapp pair generation operator in Expression
(\ref{eq:umklapp}) as
\begin{equation}
\sin(\theta_{\bf k} - 90^\circ) + \cos(\theta_{\bf k} -
90^\circ)\;b^+_{\bf k} \label{eq:umklapp-phase}
\end{equation}
It is easy to check that all the results here or in the BCS theory
will not change if we replace the $-90^\circ$ phase difference in
Eq.~(\ref{eq:umklapp-phase}) with $90^\circ$.

According to Eq.~(\ref{eq:self}) the normal pair is in a phase to
sustain superconductivity, but the umklapp pair is in a phase to
frustrate it. However, if we replace Eq.~(\ref{eq:constrain1})
with
\begin{equation}
h_{\bf k} = \frac{1}{2}\biggl[1 + \frac{\epsilon_{\bf k}}{E({\bf
k})}\biggr]\label{eq:constrain2}
\end{equation}
which can be written as
\begin{equation}
\cos^2(\theta_{\bf k} - 90^\circ) = \frac{1}{2}\biggl[1 -
\frac{\epsilon_{\bf k}}{E({\bf
k})}\biggr]\label{eq:constrain2-phase}
\end{equation}
then in Eq.~(\ref{eq:self}) $\mathcal N$ will be replaced by
$\mathcal U$ and vice versa: umklapp pairs sustain
superconductivity but normal pairs frustrate it. Apparently both
normal and umklapp pairs alone may lead to an energy gap, but
their coexistence mutually cancels their effect.

The $90^\circ$ phase difference in
Eq.~(\ref{eq:constrain1-phase}), reminiscent of the angle between
the two bars in Fig.~\ref{fig:fig3} (c), reveals an interesting
dilemma.  On the one hand the normal and umklapp pairs have to be
locked $90^\circ$ apart form each other in phase, otherwise our
trial function, Eq.~(\ref{eq:ground3}), cannot be normalized. On
the other hand, we have to pay the price that either normal or
umklapp scattering will be in a phase to frustrate
superconductivity.  Apparently only a fraction of phonons may
survive cancellation to play their traditional role in the
standard BCS theory and this, we suspect, was actually implied by
BCS some 50 years ago.

\begin{figure} \resizebox{8cm}{!}{\includegraphics{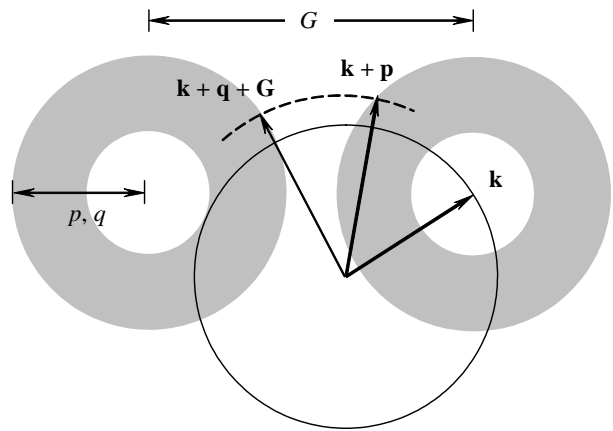}}
\caption{\label{fig:fig4} A spherical fermi surface (open circle)
and two identical phonon spheres, $G$ being the lattice constant.
Phonons in the unshaded circles do not invoke umklapp scattering.
The particle in state $\bf k$ is scattered into $\bf k + p$ and
$\bf k + q + G$ in the normal and umklapp processes, respectively,
by phonons with the same momentum.  Energy of the end state in
both processes is at the same level above the Fermi surface
(broken arc, with an exaggerated constant height above the open
circle).}
\end{figure}

\section{Average phonon frequency}
\label{sec:Average}In Eq.~(\ref{eq:self}) there can be no term in
$\bf k'(\mathcal U)$ to survive cancellation by a term in $\bf
k'(\mathcal N)$, at least for a spherical Fermi surface that does
not intersect the boundary of a Brillouin zone, and with the
constraint in Eq.~(\ref{eq:constrain1}). To see this, let $\bf k$
be a state on the Fermi surface (not in the direction of $\bf G$
for generality, see Fig.~\ref{fig:fig4}). We use $\bf k + p$ to
denote the end state of normal scattering, which is confined
within the shaded phonon sphere, of radius $p$ and centered at
$\bf k$, as a result of momentum conservation. On the other hand,
energy consideration requires that $\bf k + p$ must be at a level
near the Fermi surface, which is marked as a broken arc in
Fig.~\ref{fig:fig4}, somewhat exaggerated in its height above the
Fermi sphere.  There is an unshaded hole inside the phonon sphere
(radius $\ne q_0$, because $\bf k$ and $\bf G$ are not parallel)
where umklapp scattering is not invoked.  We assume that $\bf k +
p$ is not in the hole.

If we transport the phonon sphere to the left by a reciprocal
lattice constant $G$, then we find all the end states of umklapp
scattering allowed by momentum conservation (the hole is
completely outside the Fermi sphere but contacts its surface).
>From the intersection between the surface of the transported
phonon sphere and the broken arc in Fig.~\ref{fig:fig4}, we find a
state $\bf k + q + G$ with
\begin{equation}
q = p,\;\;\;\;\;\;\;\epsilon_{\bf k + q + G} = \epsilon_{\bf k +
p}\label{eq:p,q}
\end{equation}
i.e.\ both the phonon momentum and end state energy are the same
in both normal and umklapp scattering.  In Fig.~\ref{fig:fig4} we
have ${\bf k} = (k_x, k_y, 0)$.  When $k_z\ne 0$ we can still find
$\bf k + q + G$ to satisfy Eq.~(\ref{eq:p,q}) by moving the phonon
sphere in a proper direction over a proper distance.

We are reminded that, starting from $\bf p$, we can find more than
one $\bf q$ to satisfy Eq.~(\ref{eq:p,q}). However, we do have a
one-to-one relation between $\bf p$ and $\bf q$.  To prove this,
we reverse the process in the previous two paragraphs. We start
from $any$ end state of umklapp scattering, $\bf k + q + G$, and
then find $\bf k + p$ to satisfy Eq.~(\ref{eq:p,q}). Apparently
$\bf p$ will not fall into the unshaded hole in the phonon sphere:
this one-to-one relation is between the states in $\mathcal U$ and
those states in $\mathcal N$ which are subject to both normal and
umklapp scattering.

According to Eq.~(\ref{eq:p,q}) we have
\begin{equation}
V_{\bf k, k + q + G} =  V_{\bf k, k + p}\label{eq:V=V}
\end{equation}
where~\cite{BCS}
\begin{equation}
V_{\bf k,k + p}= \frac{2\hbar\omega_{\bf p}|\mathcal{M}_{{\bf
p}}|^2} {(\hbar\omega_{\bf p})^2 - (\epsilon_{\bf k + p} -
\epsilon_{\bf k})^2} \label{eq:V}
\end{equation}
$\mathcal M_{\bf p}$ being the matrix element, $\omega_{\bf p}$
phonon frequency, both functions of $p$ (Debye model), and $V_{\bf
k, k + q + G}$ is defined similarly. For an isotropic energy gap
we also have
\begin{equation}
\Delta({\bf k + q + G}) = \Delta({\bf k + p})\label{eq:D=D}
\end{equation}
and hence $E({\bf k + q + G}) = E({\bf k + p})$.
Eqs.~(\ref{eq:V=V}) and (\ref{eq:D=D}) lead through
Eq.~(\ref{eq:self}) to
\begin{equation}
\Delta({\bf k})= \sum_{{\bf k'}(\mathcal B)}V_{\bf k,
k'}\frac{\Delta({\bf k'})} {2E({\bf k'})} \label{eq:self3}
\end{equation}
$\mathcal B$ specifies states that do not invoke umklapp
scattering, which is confined within the unshaded circle around
$\bf k$ in Fig.~\ref{fig:fig4}.

We should remember that the Fermi sea and the reciprocal lattice
have incompatible symmetries.  Therefore in Eq.~(\ref{eq:self3})
$\mathcal B$ and hence $\Delta$ and $E$ depend on the direction of
$\bf k$ even when the Fermi sea is spherical.  For simplicity, we
have ignored the $\bf k$-dependence of $\mathcal B$, as a basis
for deriving Eq.~(\ref{eq:D=D}) from Eq.~(\ref{eq:p,q}). In future
work, we will have to generalize our proof in the previous
paragraphs by replacing Eqs.~(\ref{eq:V=V}) and (\ref{eq:D=D})
with
\begin{equation}
V_{\bf k, k + q + G}\frac{\Delta(\bf k + q + G)}{2E(\bf k + q +
G)} = V_{\bf k, k + p}\frac{\Delta(\bf k + p)}{2E(\bf k +
p)}\label{eq:VD/2E=VD/2E}
\end{equation}
as our condition to seek $\bf q$ for a given state $\bf p$ (and
vice versa). We will have to find an approximate value of $\bf q$
through the procedure in Fig.~\ref{fig:fig4}. Then we will adjust
$\bf q$ slightly (assuming that phonon states are sufficiently
numerous for small adjustment) to balance
Eq.~(\ref{eq:VD/2E=VD/2E}), which also leads to
Eq.~(\ref{eq:self3}).

Eq.~(\ref{eq:self3}) is in accord with the view that the
coexistence of normal and umklapp scattering frustrates
superconductivity. Indeed, if we cut the phonon frequency off at
the threshold of umklapp scattering, and go through the standard
BCS formalism, then we find a self-consistent equation in exactly
the same form as Eq.~(\ref{eq:self3}).

BCS introduced $\omega$, the so-called `average phonon frequency',
to define the range of phonons when they integrated the
self-consistent equation~\cite{BCS}.  In the recent literature a
liberty is often taken to replace $\omega$ with the Debye
frequency $\omega_D$. However we see from Eq.~(\ref{eq:self3})
that the BCS average phonon frequency is based on a very deep
insight. Numerically $\omega$ is likely to be significantly lower
than $\omega_D$.  According to Figs.~\ref{fig:fig1} and
\ref{fig:fig2} the phonon momentum in Eq.~(\ref{eq:self3}) cannot
exceed $q_0$ when $\bf k$ is in the direction of $\bf G$.  For a
bcc lattice with valency 1 we have $q_0/k_F = 0.267$, $k_F$ being
the radius of the Fermi sphere, or $q_0/q_D = 0.212$: when $\bf k$
is in the direction of $\bf G$ the average phonon frequency of BCS
is merely $\sim20\%$ of the Debye frequency.

\section{Excited state}
\label{sec:Excited} For the convenience of the reader, we explain
the BCS theory at finite temperatures briefly and then extend the
formulation to include umklapp scattering. BCS divided reciprocal
space into cells, $\Delta{\bf k}$, containing $N_{\bf k}$ pair
states. There are $S_{\bf k}$ single particles and $P_{\bf k}$
excited pairs in $\Delta{\bf k}$, other states being occupied by
ground pairs, giving
\begin{equation}
\frac{S_{\bf k}}{N_{\bf k}},\;\;\;\;\;\frac{P_{\bf k}}{N_{\bf
k}},\;\;\;\;\frac{N_{\bf k} - S_{\bf k} - P_{\bf k}}{N_{\bf k}}
\label{eq:Nk,Pk,Sk}
\end{equation}
as fractions of $\Delta\bf k$ occupied by single particles,
excited and ground pairs, respectively. The kinetic energy of the
particle ensemble, found from Eq.~(\ref{eq:excited}), the
electron term of Eq.~(\ref{eq:H_BCS}) and the Bloch energy contribution, is
\begin{eqnarray}
W_{\hbox{\scriptsize KE}} = \sum_{\bf k}\epsilon_{\bf
k}\Bigg[\frac{S_{\bf k}}{N_{\bf k}} + 2(1 - h_{\bf k})\frac{P_{\bf
k}}{N_{\bf k}}\;\;\;\;\;\;\;\;\;\;\;\;\;\nonumber\\
+ 2h_{\bf k}\frac{N_{\bf k} - S_{\bf k} - P_{\bf k}}{N_{\bf
k}}\Bigg]\label{eq:W_KE}
\end{eqnarray}
or \begin{equation} W_{\hbox{\scriptsize KE}}= 2\sum_{\bf
k}\epsilon_{\bf k}\Big[f_{\bf k} + (1 - 2f_{\bf k})h_{\bf
k}\Big]\label{eq:W_KE2}
\end{equation}
with
\begin{equation}
f_{\bf k} = \frac{S_{\bf k}}{2N_{\bf k}} + \frac{P_{\bf k}}{N_{\bf
k}}\label{eq:f_k}
\end{equation}
In Eqs.~(\ref{eq:W_KE}) and (\ref{eq:W_KE2}) $\epsilon_{\bf k}$ is
averaged in $\Delta{\bf k}$ (a cell), $\bf k$ runs over the cells
(one $\bf k$ for each cell). The interaction energy is found to be
\begin{eqnarray}
W_{\hbox{\scriptsize I}} = -\sum_{\bf k, k'}V_{\bf k,
k'}\sqrt{h_{\bf k}(1 - h_{\bf k})}(1 - 2f_{\bf k})\;\nonumber\\
\times\sqrt{h_{\bf k'}(1 - h_{\bf k'})}(1 - 2f_{\bf k'})
\label{eq:W_I}
\end{eqnarray}
The entropy is given by
\begin{equation}
S = -2k_B\sum_{\bf k}\Big[f_{\bf k}\ln f_{\bf k} + (1 - f_{\bf
k})\ln(1 - f_{\bf k})\Big] \label{eq:S}
\end{equation}
$k_B$ being the Boltzmann constant, which leads to the free energy
\begin{equation}
F = W_{\hbox{\scriptsize KE}} + W_{\hbox{\scriptsize I}} - TS
\label{eq:F}
\end{equation}
We have
\begin{eqnarray}
&&\frac{\partial F}{\partial h_{\bf k}} = 2\epsilon_{\bf k}(1 -
2f_{\bf k}) - \frac{1 - 2h_{\bf k}}{\sqrt{h_{\bf k}(1 - h_{\bf
k})}}(1 - 2f_{\bf k})\nonumber\\
&&\;\;\;\;\;\;\;\times\;\sum_{\bf k'}V_{\bf k, k'}\sqrt{h_{\bf
k'}(1 - h_{\bf k'})}\;(1 - 2f_{\bf k'})\label{eq:dF/dh}
\end{eqnarray}
which can be compared with Eq.~(\ref{eq:dW1}). When $\partial
F/\partial h_{\bf k} = 0$, Eq.~(\ref{eq:dF/dh}) leads through
Eq.~(\ref{eq:constrain1}) to
\begin{equation}
\Delta({\bf k})= \sum_{\bf k'}V_{\bf k, k'}\frac{\Delta({\bf k'})}
{2E({\bf k'})}(1 - 2f_{\bf k'}) \label{eq:self4}
\end{equation}
It is interesting that $\bf k'$ runs over the same range at any
$T$. Therefore, when the temperature rises, ground pairs are
replaced by single particles and excited pairs gradually in each
and every cell, $\Delta\bf k'$, rather than being driven out
altogether cell by cell. We also have
\begin{equation}
\frac{\partial F}{\partial f_{\bf k}} = 2E({\bf k}) +
2k_BT\ln\frac{f_{\bf k}}{1 - f_{\bf k}}\label{eq:dF/df}
\end{equation}
or
\begin{equation}
1 - 2f_{\bf k} = \tanh\frac{E({\bf k})}{2k_BT}\label{eq:1-2f}
\end{equation}
when $\partial F/\partial f_{\bf k} = 0$.  Combining
Eqs.~(\ref{eq:self4}) and (\ref{eq:1-2f}), we find
\begin{equation}
\Delta({\bf k})= \sum_{\bf k'}V_{\bf k, k'}\frac{\Delta({\bf k'})}
{2E({\bf k'})}\tanh\frac{E({\bf k'})}{2k_BT} \label{eq:self5}
\end{equation}
which is the self-consistent equation at $T > 0$ in the BCS
theory~\cite{BCS}.

Eq.~(\ref{eq:self5}), though sophisticated and elegant, manifests
a simple idea, viz\ all states fall into two categories: those
allowed at $T = 0$ and those not, to evaluate the entropy in
Eq.~(\ref{eq:S}). Once we endorse this idea, the rest of the
derivation is automatic. For each $\bf k$ we have three
configurations, whose contributions to the ensemble energy are
well-defined. Thermodynamics governs the number of the particles
in each of the configurations, which in turn governs the magnitude
of the energy gap. Apparently, this simple idea will not change
when umklapp scattering is included.

In the presence of umklapp scattering, we replace
Eq.~(\ref{eq:excited}) with:
\begin{eqnarray}
&& |\Psi\rangle_{\mbox{exc}} = \prod _{{\bf k''}(\mathcal
S_1)}\prod _{{\bf k''}(\mathcal S_2)}a^+_{{\bf k''}\sigma}\prod
_{{\bf k'}(\mathcal P_1)}\prod_{{\bf
k'}(\mathcal{G}_2)}\big(\sqrt{1-h_{\bf k'}}\;b^+_{\bf k'}\nonumber\\
&& \;\;\;\;- \sqrt{h_{\bf k'}}\big)\prod _{{\bf k}({\mathcal
G}_1)}\prod _{{\bf k}({\mathcal P}_2)}\big(\sqrt{1-h_{\bf k}} +
\sqrt{h_{\bf k}}\;b^+_{\bf k}\big)|0\rangle \label{eq:excited2}
\end{eqnarray}
where $\mathcal G$ and $\mathcal P$ specify ground and excited
pairs, respectively, $\mathcal S$ specifies single particles, as
in Eq.~(\ref{eq:excited}), indices 1 and 2 specify the normal and
umklapp processes, i.e.\ we separate $\mathcal N$ and $\mathcal U$
into $\mathcal S_1$, $\mathcal P_1$, $\mathcal G_1$ and $\mathcal
S_2$, $\mathcal P_2$, $\mathcal G_2$ respectively. Note that, when
ground pairs are generated by Expression (\ref{eq:normal}), we use
Expression (\ref{eq:umklapp}) to generate excited pairs, similar
to the arrangement in Eq.~(\ref{eq:excited}).  On the other hand,
when ground pairs are generated by Expression (\ref{eq:umklapp}),
we use Expression (\ref{eq:normal}) to generate excited pairs,
which raises the energy of the pair ensemble, in recognition that
umklapp pairs make less negative contribution to that energy the
fewer their number.

We divide $\mathcal N$ and $\mathcal U$ into cells $\Delta{\bf
k}_1$ and $\Delta{\bf k}_2$ respectively.  We have
\begin{equation}
\frac{S_{1\bf k}}{N_{1\bf k}},\;\;\;\;\;\frac{P_{1\bf k}}{N_{1\bf
k}},\;\;\;\;\frac{N_{1\bf k} - S_{1\bf k} - P_{1\bf k}}{N_{1\bf
k}} \label{eq:N1k,P1k,S1k}
\end{equation}
as fractions of $\Delta{\bf k}_1$ occupied by single particles,
excited and ground pairs.  We also have
\begin{equation}
\frac{S_{2\bf k}}{N_{2\bf k}},\;\;\;\;\;\frac{P_{2\bf k}}{N_{2\bf
k}},\;\;\;\;\frac{N_{2\bf k} - S_{2\bf k} - P_{2\bf k}}{N_{2\bf
k}} \label{eq:N2k,P2k,S2k}
\end{equation}
as fractions of $\Delta{\bf k}_2$ occupied by single particles,
excited and ground pairs. Letting
\begin{equation}
f_{\bf k} = \frac{S_{1\bf k}}{2N_{1\bf k}} + \frac{P_{1\bf
k}}{N_{1\bf k}}\;\;\;\;\hbox{or}\;\;\;\;\frac{S_{2\bf k}}{2N_{2\bf
k}} + \frac{P_{2\bf k}}{N_{2\bf k}}\label{eq:f2_k}
\end{equation}
for $\bf k$ specified by $\mathcal N$ or $\mathcal U$
respectively, we find
\begin{eqnarray}
W_{\hbox{\scriptsize KE}} &=& \sum_{\bf k(\mathcal
N)}2\epsilon_{\bf k}\bigl[f_{\bf k} + (1 - 2f_{\bf k})h_{\bf
k}\bigr]\nonumber\\
&+& \sum_{\bf k(\mathcal U)}2\epsilon_{\bf k}\bigl[f_{\bf k} + (1
- 2f_{\bf k})(1 - h_{\bf k})\bigr]\label{eq:W_KE3}
\end{eqnarray}
as the kinetic energy, which can be compared with
Eq.~(\ref{eq:W_KE2}).  We also find
\begin{equation}
W_{\hbox{\scriptsize I}}= -W_{11} - W_{22} + W_{12} +
W_{21}\label{eq:W_I2}
\end{equation}
with
\begin{eqnarray}
W_{11} = \sum_{\bf k(\mathcal N)}\sum_{\bf k'(\mathcal N)}V_{\bf
k, k'}\sqrt{h_{\bf k}(1 - h_{\bf
k})}\sqrt{h_{\bf k'}(1 - h_{\bf k'})}\nonumber\\
\times(1 - 2f_{\bf k})(1 - 2f_{\bf k'})\nonumber
\end{eqnarray}
and
\begin{eqnarray} W_{12} =\sum_{\bf k(\mathcal N)}\sum_{\bf
k'(\mathcal U)}V_{\bf k, k'}\sqrt{h_{\bf k}(1 - h_{\bf
k})}\sqrt{h_{\bf k'}(1 - h_{\bf k'})}\nonumber\\
\times(1 - 2f_{\bf k})(1 - 2f_{\bf k'})\nonumber
\end{eqnarray}
$W_{21}$ and $W_{22}$ are identical to $W_{12}$ and $W_{11}$,
respectively, save that $\mathcal N$ is replaced by $\mathcal U$
and vice versa.

Now we minimize the free energy $F$, which is defined by
Eqs.~(\ref{eq:F}), (\ref{eq:W_KE3}) and (\ref{eq:W_I2}) when $\bf
k$ is specified by $\mathcal N$. We find
\begin{eqnarray}
\frac{\partial F}{\partial h_{\bf k}} = 2\epsilon_{\bf k}(1 -
2f_{\bf k}) - \frac{1 - 2h_{\bf k}}{\sqrt{h_{\bf k}(1 - h_{\bf
k})}}(1 - 2f_{\bf k})\nonumber\\
\times\biggl[\sum_{\bf k'(\mathcal N)} - \sum_{\bf k'(\mathcal
U)}\biggr]V_{\bf k, k'}\sqrt{h_{\bf k'}(1 - h_{\bf k'})}\;(1 -
2f_{\bf k'})\label{eq:dF2/dh}
\end{eqnarray}
which can be compared with Eq.~(\ref{eq:dF/dh}).  When $\partial
F/\partial h_{\bf k} = 0$, Eq.~(\ref{eq:dF/dh}) leads through
Eq.~(\ref{eq:constrain1}) to
\begin{equation}
\Delta({\bf k})= \biggl[\sum_{\bf k'(\mathcal N)}-\sum_{\bf
k'(\mathcal U)}\biggr]V_{\bf k, k'}\frac{\Delta({\bf k'})}
{2E({\bf k'})}(1 - 2f_{\bf k'}) \label{eq:self6}
\end{equation}
which can be compared with Eq.~(\ref{eq:self4}).
Eqs.~(\ref{eq:dF/df}) and (\ref{eq:1-2f}) do not change in the
presence of umklapp scattering, and this leads through
Eq.~(\ref{eq:self6}) to
\begin{equation}
\Delta({\bf k})= \Biggl[\sum_{\bf k'(\mathcal N)} - \sum_{\bf
k'(\mathcal U)}\Biggr]V_{\bf k, k'}\frac{\Delta({\bf k'})}
{2E({\bf k'})}\tanh\frac{E({\bf k'})}{2k_BT} \label{eq:self7}
\end{equation}
as the self-consistent equation for $T > 0$ in the presence of
umklapp scattering.  Eq.~(\ref{eq:self7}) can be compared with
Eq.~(\ref{eq:self5}).

When $\bf k$ is specified by $\mathcal U$ we define the free
energy as
\begin{equation}
F = -W_{\hbox{\scriptsize KE}} - W_{\hbox{\scriptsize I}} - TS
\label{eq:F2}
\end{equation}
which can be compared with Eq.~(\ref{eq:F}), $W_{\hbox{\scriptsize
KE}}$ and $W_{\hbox{\scriptsize I}}$ are defined in
Eqs.~(\ref{eq:W_KE3}) and (\ref{eq:W_I2}) respectively, $S$ is
defined in Eq.~(\ref{eq:S}). In Eq.~(\ref{eq:F2}) $W =
W_{\hbox{\scriptsize KE}} + W_{\hbox{\scriptsize I}}$ is negative,
because umklapp pairs make negative contributions to the energy of
the pair ensemble.  This means that we have $\delta Q = -dW$ if
the change in $Q$ is caused by the change in the number and/or
occupancy of umklapp pairs, $Q$ stands for heat. Therefor we have
to let $F = -W - TS$, in order to have $dF = \delta Q - TdS - SdT
= -SdT$, which is the correct expression of $dF$ (for constant
volume). Now Eq.~(\ref{eq:dF2/dh}) becomes
\begin{eqnarray}
\frac{\partial F}{\partial h_{\bf k}} = -2\epsilon_{\bf k}(1 -
2f_{\bf k}) + \frac{1 - 2h_{\bf k}}{\sqrt{h_{\bf k}(1 - h_{\bf
k})}}(1 - 2f_{\bf k})\nonumber\\
\times\biggl[\sum_{\bf k'(\mathcal N)} - \sum_{\bf k'(\mathcal
U)}\biggr]V_{\bf k, k'}\sqrt{h_{\bf k'}(1 - h_{\bf k'})}\;(1 -
2f_{\bf k'})\label{eq:dF3/dh}
\end{eqnarray}
which also leads to Eq.~(\ref{eq:self6}).  We still have
Eqs.~(\ref{eq:dF/df}) and (\ref{eq:1-2f}), which lead through
Eq.~(\ref{eq:self6}) to Eq.~(\ref{eq:self7}).

If we adopt the view that superconductivity is frustrated when
normal and umklapp scattering coexist, then we can go through the
standard BCS formalism at $T > 0$ to find
\begin{equation}
\Delta({\bf k})= \sum_{\bf k'(\mathcal B)}V_{\bf k,
k'}\frac{\Delta({\bf k'})} {2E({\bf k'})}\tanh\frac{E({\bf
k'})}{2k_BT}\label{eq:self8}
\end{equation}
which is identical to Eq.~(\ref{eq:self7}), $\mathcal B$ (defined
in Section~\ref{sec:Average}) specifies the range of $\bf k'$,
which does not depend on $T$, similar to the range of $\bf k'$ in
Eq.~(\ref{eq:self4}) or (\ref{eq:self5}).

\section{Conclusions}
\label{sec:Conclusions}In the conventional picture of single
electron scattering in a metal the normal and umklapp scattering
processes are associated with distinctively different ranges of
momentum transfer. There is no possibility for these two processes
to interfere with each other in the sense that, with the same
initial state, the end states of normal and umklapp scattering do
not coincide.  In this picture umklapp scattering is effectively
treated as normal scattering in an extended range of momentum
transfer and, when applied in the BCS theory, it tells us that
umklapp scattering also lowers the ensemble energy and thus
enhances superconductivity~\cite{Pines, Semenov}.

We recognize that, when the scattered entity is not a single
electron but a pair, the end states of normal and umklapp
scattering may coincide and the two processes may interfere. We
find a formal deficiency in the standard BCS ground state wave
function, which is designed to count the statistics of scattering
of pairs in the normal or umklapp process alone and leads to a
self-contradiction when these two processes coexist.  A suitable
trial function locks the end states of normal and umklapp
scattering 90$^\circ$ apart in phase but leads to the consequence
that superconductivity is frustrated when the normal and umklapp
processes coexist. Then only a fraction of the phonons actually
takes part in electron pairing. Our conclusion is reached within
the BCS formalism which is based on the variational principle.

Another approach, the Eliashberg formalism, is based on the
one-electron Green's function, and here too superconductivity is
enhanced when umklapp scattering is introduced~\cite{Carbotte}.
However, in this formalism, which is non-variational and does not
involve a trial function, the physics of pair scattering and
interference effects may be manifest in a manner different from
that in the BCS variational approach. Further consideration of the
contribution of umklapp phonons in different formalisms would
appear to be merited.

\end{document}